\documentclass{article}
\usepackage[final]{neurips_2020}
\usepackage{subcaption}
\usepackage{amsmath,amssymb,amsfonts}
\usepackage{algorithmic}
\usepackage{url}
\usepackage{textcomp}
\usepackage{changepage}
\usepackage[utf8]{inputenc} 
\usepackage[T1]{fontenc}    
\usepackage{hyperref}       
\usepackage{url}            
\usepackage{booktabs}       
\usepackage{nicefrac}       
\usepackage{microtype}      
\usepackage{lipsum}
\usepackage{multirow}
\usepackage[table,xcdraw]{xcolor}
\usepackage{graphicx}
\usepackage{comment}
\usepackage{caption}
\usepackage{subcaption}
\usepackage{floatrow}
\newfloatcommand{capbtabbox}{table}[][\FBwidth]

\begin{document}

\title{Proximity Sensing: Modeling and Understanding Noisy RSSI-BLE Signals and Other Mobile Sensor Data for Digital Contact Tracing}
\author{Sheshank Shankar\And Rishank Kanaparti\And Ayush Chopra\And Rohan Sukumaran\And Parth Patwa\And Myungsun Kang\And Abhishek Singh\And Kevin P. McPherson\And Ramesh Raskar\\\\PathCheck Foundation \quad MIT Media Lab\\Cambridge, MA\\\texttt{sheshank.shankar@pathcheck.org}}

\date{September 2020}
\maketitle

\begin{abstract}
As we await a vaccine, social-distancing via efficient contact tracing has emerged as the primary health strategy to dampen the spread of COVID-19. To enable efficient digital contact tracing, we present a novel system to estimate pair-wise individual proximity, via a joint model of Bluetooth Low Energy (BLE) signals with other on-device sensors (accelerometer, magnetometer, gyroscope). We explore multiple ways of interpreting the sensor data stream (time-series, histogram, etc) and use several statistical and deep learning methods to learn representations for sensing proximity. We report the normalized Decision Cost Function (\textit{nDCF}) metric and analyze the differential impact of the various input signals, as well as discuss various challenges associated with this task. 

\end{abstract}
\section{Introduction}
As economies open up, digital contact tracing is emerging as an important tool to help contain the spread of COVID-19 by providing exposure notification to individuals who come in close proximity to infected individuals \cite{ cheng2020contact, ferretti2020quantifying, macintyre2020case, zhang2020changes}. There have been several proposals for proximity sensing varying across different modalities, including WiFi signals \cite{dmitrienko2020proximity, Sapiezynski, trivedi}, GPS \cite{raskar2020adding}, Ultrasound \cite{novid}, and QR codes \cite{QR}. However, Bluetooth is the most widely accepted technology for digital contact tracing, as it is the most feasible across the available options. Moreover, it is supported by the Google Apple Exposure Notification (GAEN) API \cite{gaen}. 
Current approaches for automated exposure notification use BLE signals emanating from smartphones (chirps) to detect if a person has been in close proximity to an infected individual. However, there are certain limitations with the accuracy of Bluetooth, as described in \cite{leith2020coronavirus, leith2020measurement, Contact-tracing}. Researchers have tried augmenting Bluetooth-based protocols with other modalities (such as Ultrasound \cite{novid} and GPS \cite{raskar2020adding}), however, that does not circumvent the signal processing limitation present in Bluetooth based systems. Existing attempts at calibrating Bluetooth data only use device level information \cite{GoogleEN} to partially account for hardware level differences among the devices. In addition, the received signal strength indicator (RSSI) value of Bluetooth chirps sent between phones is a noisy estimator of the actual distance between the phones as they can be dramatically affected by real-world conditions \cite{hatke2020using}. 

In this paper, we predict the distance between two phones using the RSSI values from BLE signal logs, along with data from other mobile sensors (as described in Table \ref{fig:sen_des}). The additional sensors help account for the real-world complexities that affect the variations in BLE signals, such as the position of the phone, movement, etc. Using datasets collected by NIST and MITRE, we experiment with various machine learning models to learn the representations of the data, achieving the most favorable results by using a temporal Conv1D. Additionally, to understand the contribution of various sensors on the overall task as well as the extent of noise present in this data distribution, we perform ablation studies and discuss various challenges associated with the data distribution.
\section{Preliminaries}

In our experiments, we use the NIST dataset (collected in coordination with the MIT PACT project) and the MITRE Range Angle Structured (further referred to as MITRE) dataset. Both the datasets are collected using the Structured Contact Tracing Protocol V 2.0~\cite{dataset_protocol} and provide a stream of Bluetooth RSSI data between two phones at the following distances, 1.2, 1.8, 3.0, and 4.5 meters. We use these distances as classes. 
In addition, the dataset collects other phone sensor data (listed in Table \ref{fig:sen_des}), as well as experiment level metadata (such as device models, power, carriage states, etc). 

The dataset is formatted as multiple-segmented intervals of continuous sensor readings (between transmitter and receiver) at the same distance. This is done for multiple 4-second device interactions per experiment. We primarily use the MITRE and a set-aside subset of the NIST datasets for training and evaluation respectively. However, results of models trained on a separate subset of the NIST dataset are also discussed. 

\section{Methodology}
\label{method}

\begin{figure}[h!]
\includegraphics[width=10cm]{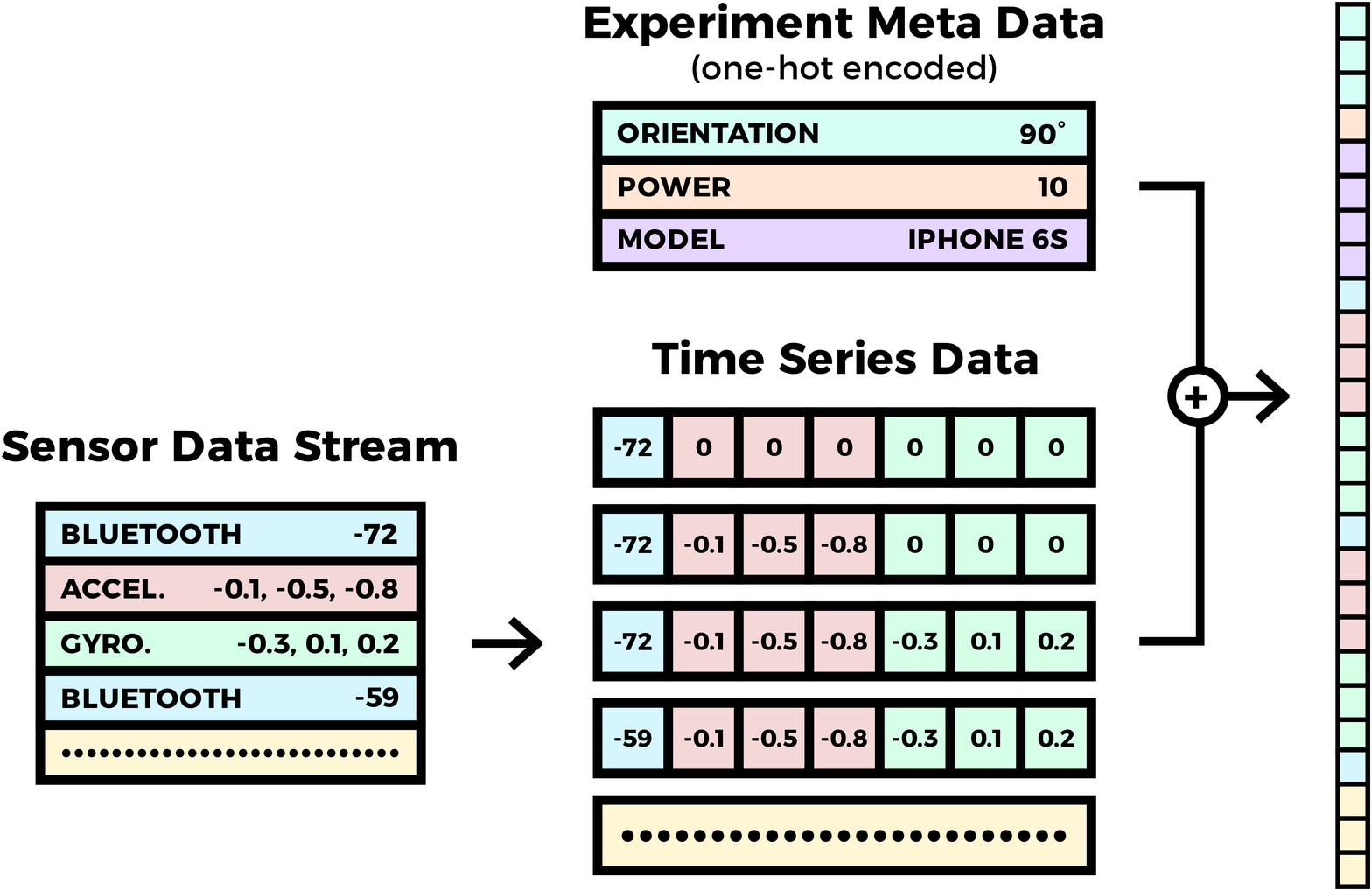}
\caption{A \textit{simplified} visualization of the data processing pipeline for the \textbf{concatenated} time-steps input. After the sensor data stream is converted into a time series of 150 timesteps, they are concatenated and merged with the one-hot encoded meta-data to create a single feature vector.}
\label{data-processing}
\end{figure}

As can be seen in Figure \ref{fig:pca}, the raw Bluetooth RSSI and other sensor data is extremely noisy. Hence, we exploit the temporal characteristics of the dataset by modeling each experiment as a time series, breaking each 4-second interval into multiple time-steps. To minimize noise and the need for over-sampling and under-sampling readings, we choose the mean number of samples per each 4-second interval (150) as the fixed length of each time series. Every time-step is represented as a normalized fixed-length feature vector representing the most recent values obtained from each sensor. In addition, the metadata is one-hot encoded and concatenated to each timestep's vector. Figure \ref{data-processing} shows the data processing pipeline for the non-temporal models that do not make use of a time-series input.

Inspired by \cite{Tianlang}, we also experiment with a histogram representation of the BLE signals. Here, we use the frequencies of RSSI values in various buckets instead of using the actual RSSI values. We also try using mix-up data augmentation \cite{zhangMixup}, to reduce overfitting by increasing the variations present in the MITRE dataset. However, it did not provide a significant performance increase.


To model this complex data, we use classifiers derived from various DL models (feed-forward, LSTM, GRU, temporal one-dimensional convolutional network, and a convolutional GRU \cite{convgru}), Decision Tree-based models (XGBoost \cite{chen2016xgboost}, Random Forests \cite{breiman2001random}), as well as with support vector machines (SVM) \cite{cortesSVM} and a Naive Bayes Classifier. The majority of the models are trained on the entire MITRE dataset, however, the SVM and tree-based models are only trained on a smaller subset of the dataset. This is because training on the full dataset for these model architectures do not complete in a reasonable amount of time. To understand the variations caused by minor changes in the input, we also model the task as a regression problem, where we round the outputs to the nearest class. 

We evaluate results on a test set from MITRE and NIST datasets with the \textit{normalized Decision Cost Function (nDCF)} metric \cite{TC4TL} that effectively combines the probability of a false negative and false positive into a single value using weights reflecting the relative cost of each type of error. 

All experiments were run on a Intel(R) Xeon(R) CPU E5-2650 v4 @ 2.20GHz server with 528 GB RAM, 48 cores, and on a single Nvidia 1080 Ti GPU. The temporal networks are built using PyTorch \cite{NEURIPS2019_9015}, whereas the Support Vector Machine and the Decision Tree based models are implemented using scikit-learn \cite{pedregosa2011scikit}. All  the deep learning experiments use the Adam optimizer \cite{kingma2014adam} and optimize Categorical Crossentropy loss. We plan to open source the code.
\section{Results and Analysis}

After rigorous parameter tuning, we present the best results on various model architectures evaluated on a subset of the NIST dataset and trained on either the MITRE dataset or a separate subset of the NIST dataset. As highlighted in Table~\ref{fig:ndcf_results}, the temporal one-dimensional convolutional network shows the best performance. Surprisingly, the Nu-Support Vector Classifier is able to perform relatively well even with a substantially smaller training subset of the MITRE dataset (as described in the~\nameref{method} section). This suggests that there could be less variation in the MITRE dataset, contributing to lower levels of generalization. Additionally, for the random forest model architecture, using a regressor instead of a classifier decreases the nDCF by 0.19, possibly indicating that minor variations in the input could be relevant. 
The Convolutional GRU architecture performs extremely well when trained on the training subset of the NIST dataset, but significantly loses performance when trained on the MITRE dataset. Due to the complexity of the Convolutional GRU, we hypothesize that this model is overfitting to the conditions of the \textit{location} that the NIST data was collected in. 

\begin{figure}[!h]
\begin{floatrow}
\RawFloats
\capbtabbox{%
\begin{tabular}{|l|l|}
\hline
\textbf{Name} & \textbf{Description}                                                        \\ \hline
Bluetooth       & BLE  strength                                                            \\ \hline
Accelerometer   & linear acceleration                                                      \\ \hline
Gyroscope       & angular velocity                                                         \\ \hline
Magnetometer    & \begin{tabular}[c]{@{}l@{}}magnetic \\ aberration\\ changes\end{tabular} \\ \hline
Attitude        & roll, pitch, yaw                                                         \\ \hline
Gravity         & \begin{tabular}[c]{@{}l@{}}Gravity along \\ X, Y, Z axes\end{tabular}    \\ \hline
Altitude        & \begin{tabular}[c]{@{}l@{}}atmospheric\\ pressure changes\end{tabular}   \\ \hline
Compass         & \begin{tabular}[c]{@{}l@{}}orientation \&\\  direction\end{tabular}      \\ \hline
\end{tabular}}{
\caption{Descriptions of mobile sensors present in both datasets.}
\label{fig:sen_des} }

\RawFloats
\capbtabbox{%
\begin{tabular}{|l|l|l|l|l|l|}
\hline
\textbf{Network} & \textbf{Train Set} & \textbf{nDCF} \\ \hline
ConvGRU & MITRE & 1.02 \\ \hline
RNN (GRU) & MITRE & 0.97 \\ \hline
Feed Forward & MITRE & 0.77 \\ \hline
\textbf{Conv1D} & \textbf{MITRE} & \textbf{0.58} \\ \hline
Conv1D (MaxPool) & MITRE & 0.81 \\ \hline
Conv1D (Dilation) & MITRE & 0.70 \\ \hline
C-SVC & MITRE & 0.99 \\ \hline
Nu-SVC & MITRE & 0.77 \\ \hline
XGBoost & MITRE & 1.02 \\ \hline
RF Classifier & MITRE & 1.04 \\ \hline
RF Regressor & MITRE & 0.85 \\ \hline
RF Histogram (Regressor) & MITRE & 0.90 \\ \hline
Naive Bayes & MITRE & 0.92 \\ \hline
GRU & NIST & 0.28 \\ \hline
\textbf{ConvGRU} & \textbf{NIST} & \textbf{0.16} \\ \hline
\end{tabular}}{%
\caption{Results for various model representations, evaluated on a subset of the MITRE and NIST dataset.}
\label{fig:ndcf_results}
}
\end{floatrow}
\end{figure}

\subsection{Ablation studies}
We perform preliminary ablation studies by excluding data from input data streams to estimate the role of different sensors in proximity sensing. 
 Due to the extensive number of potential subsets, we limit our experiments to only subsets based on a physics intuition. For example, we find using only the RSSI-BLE readings increases the divergence in the final training and testing accuracy, indicating higher susceptibility to overfitting. After training with multiple other combinations of sensors, we achieve the best performance using the gyroscope, accelerometer, magnetometer, and the RSSI-BLE readings. When excluding experiment-level metadata, we do not observe any significant performance change. We find that training is actually susceptible to overfitting on two classes when we include the experiment-level metadata. 




%



\section{Discussion}
\label{discussion}
\begin{figure}
\centering
\begin{subfigure}{.5\textwidth}
  \centering
  \includegraphics[width=.8\linewidth]{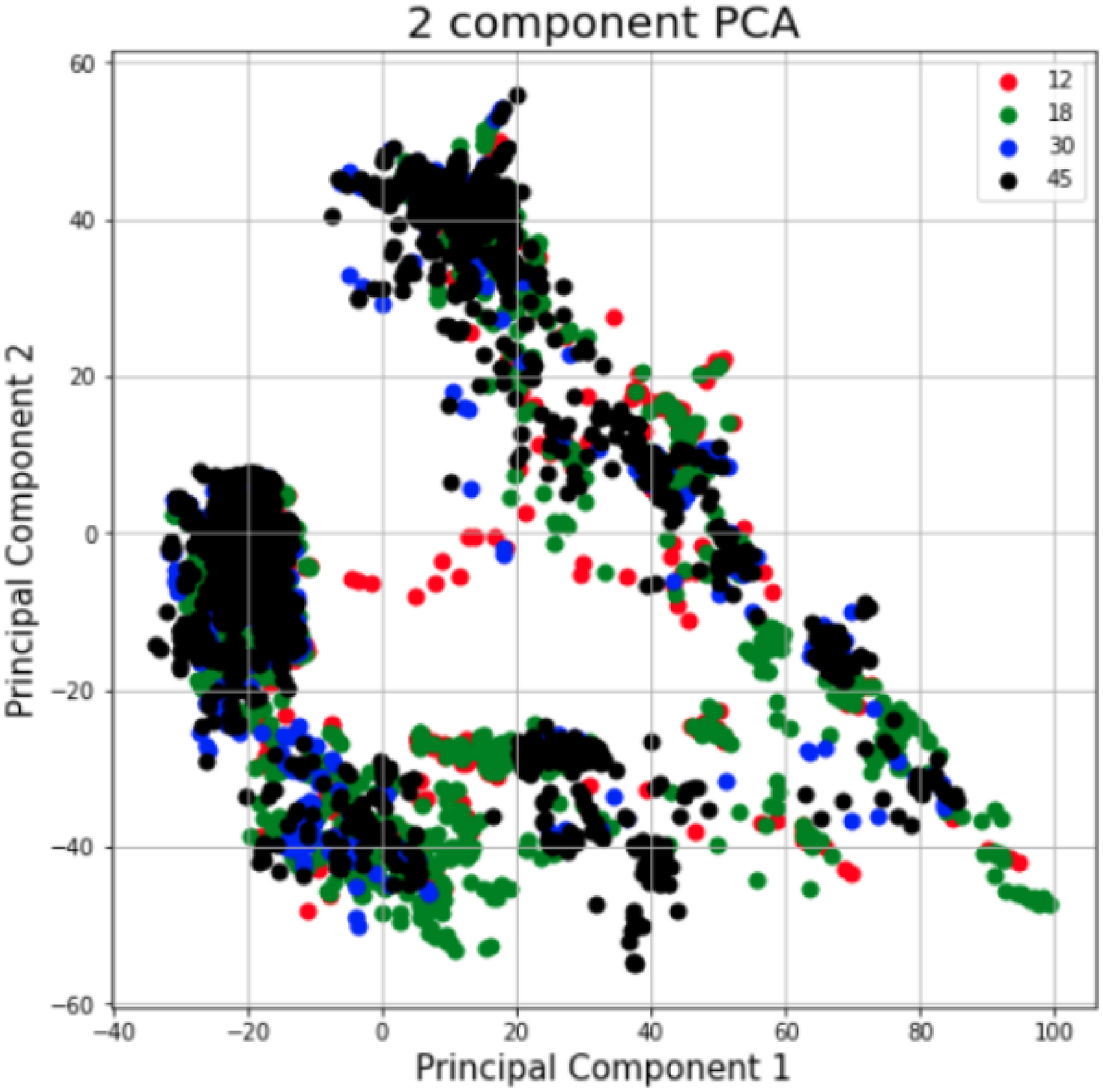}
  \label{fig:sub1}
\end{subfigure}%
\begin{subfigure}{.5\textwidth}
  \centering
  \includegraphics[width=.8\linewidth]{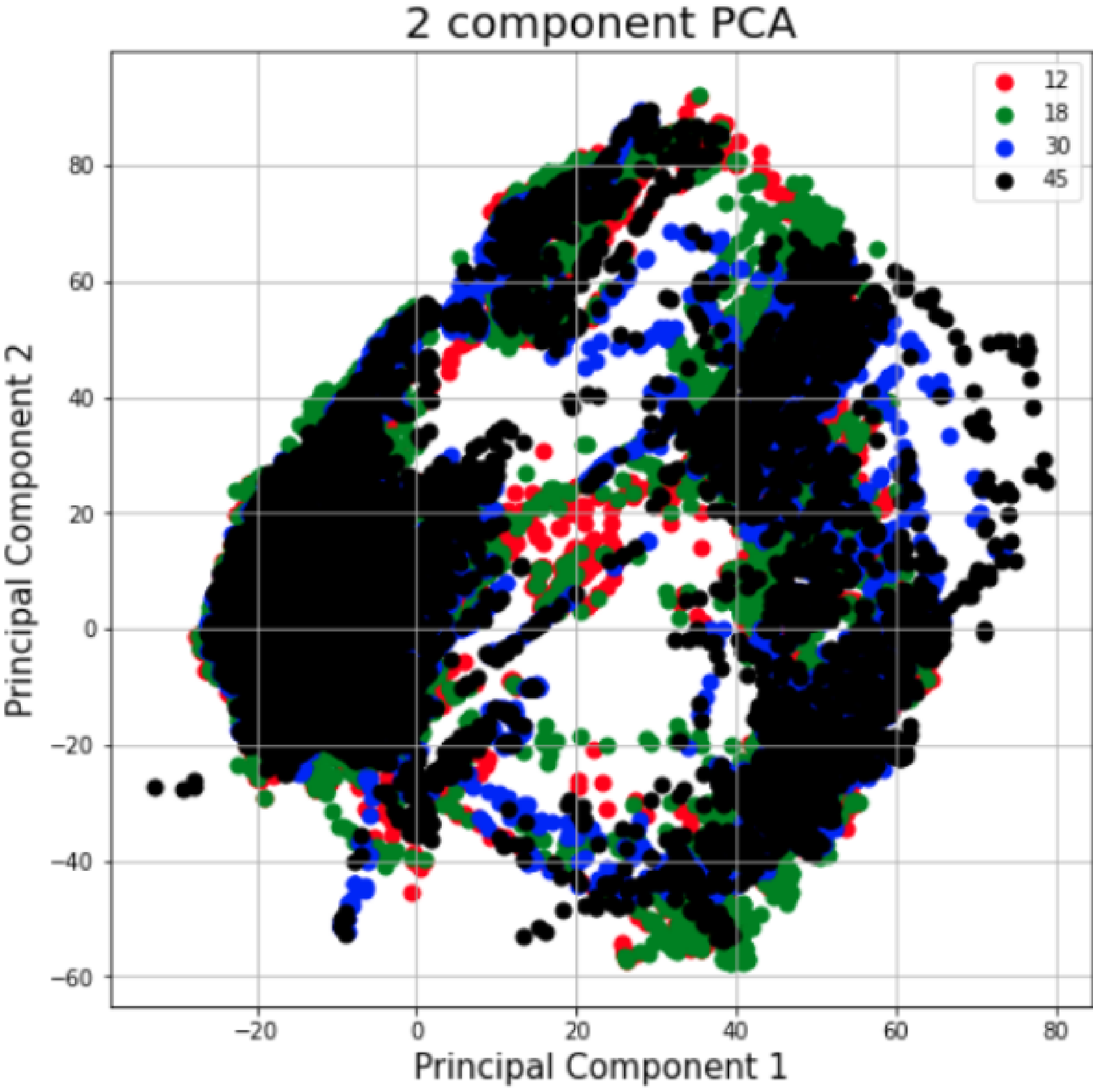}
  \label{fig:sub2}
\end{subfigure}
\caption{The PCA of the MITRE (left) \& NIST (right) datasets shows there are no clear decision boundaries present among the classes in this low dimensional representation.}
\label{fig:pca}
\end{figure}

In this section, we discuss the feasibility of proximity sensing with respect to the distribution of data. We perform low dimensional projections of the dataset to identify underlying target class clusters in the feature space. Figure~\ref{fig:pca} shows the Principal Component Analysis (PCA) plots for both datasets. In both plots, clusters are heavily overlapping, without a clear decision boundary for any two labels. This could be caused by external factors such as, physical barriers present between phones, the number and time spread of observed chirps, and multi-path signals reflected from surfaces (e.g. indoor vs outdoor). However, higher dimensional hyper-planes dividing the classes may exist.


A major challenge we encounter across experiments is the lack of generalization of the models trained on MITRE dataset to the NIST dataset. To assess the generalizability of the NIST dataset, we train on NIST dataset and evaluation on the MITRE dataset, but it does not yield any notable improvement in generalization. Informed by these results, we try to estimate the gap between the MITRE and NIST distributions. The inconsistencies between the two distributions can be observed by the high values of $\ell_2$ between the pair-wise distances of any two feature vectors. Looking at the cross-dataset nearest neighbour pairs, we find that a significant number of the pairs had different classes. Further proving this data discrepancy, we receive similar results when training on an \textit{optimal training subset} that includes only the $2$ nearest points in the MITRE dataset for each point in the NIST dataset is included. We also find that the average distance between a nearest-neighbor pair have an $\ell_2$ distance of $24$; however, if we isolate nearest-neighbor pairs with different classes, the average $\ell_2$ distance is around $200$. This supports the argument that the training and evaluation data distributions are not similar enough to capture any generalizable information. However, a thorough statistical analysis is needed to confirm this argument, as highly non-linear manifolds that can fit both distributions may exist.


\section{Conclusion and Future Work}
In this paper, we present our approach for detecting proximity between phones for digital contact tracing. Compared to current systems that only rely on RSSI values, our approach attempts to fuse knowledge from additional sensors available on smartphones to mitigate the noise present in the Bluetooth signal. Out of all the models we try, Conv1D performs the best. We report our findings and analysis over different data streams and methods. Effectively sensing proximity was marked by several challenges due to the noise in the data distribution and hence the lack of generalization. 

In the future, we plan to work on more interpretable modeling and extensive breakdown of different sensor contribution to the prediction. A combination of physics based forward model and data driven predictor will also be a good step towards robustly detect proximity. We plan to incorporate other co-location technologies (such as WiFi, GPS, Ultrasound, and QR) with our current approach to provide more accurate results.

\bibliographystyle{aaai21}
\bibliography{references}
\end{document}